\documentclass[longauth]{aa} % for the long lists of affiliations 
\usepackage{graphicx}
\usepackage{txfonts}
\usepackage{natbib,twoopt}
\usepackage{hyperref} %% to avoid \citeads line fills
\bibpunct{(}{)}{;}{a}{}{,}             %% natbib format for A&A and ApJ
\begin{document} 

\title{The \emph{Gaia}-ESO Survey: Inhibited extra mixing in two giants of the open cluster Trumpler 20?\thanks{Based on observations collected at the European Organisation for Astronomical Research in the Southern Hemisphere under ESO programme 188.B-3002 (The Gaia-ESO Public Spectroscopic Survey).}}

\author{R. Smiljanic\inst{1}
          \and
         E. Franciosini\inst{2}
          \and
         S. Randich\inst{2}
         \and
         L. Magrini\inst{2}
         \and
         A. Bragaglia\inst{3}
         \and
         L. Pasquini\inst{4}
         \and
         A. Vallenari\inst{5}
         \and
         G. Tautvai{\v s}ien{\.e}\inst{6}
         \and
         K. Biazzo\inst{7}
         \and
         A. Frasca\inst{7}
         \and
         P. Donati\inst{3,8}
         \and
         E. Delgado Mena\inst{9}
         \and
         A.~R. Casey\inst{10} 
          \and
          D. Geisler\inst{11}
          \and
          S. Villanova\inst{11}
         \and
         B. Tang\inst{11} 
         \and
         S. ~G. Sousa\inst{9}
         \and
         G. Gilmore\inst{10}
         \and
         T. Bensby\inst{12}
         \and
         P. Fran\c cois\inst{13,14}
         \and
         S.~E. Koposov\inst{10}
         \and
         A.~C. Lanzafame\inst{15}
         \and
         E. Pancino\inst{2,16}
         \and
         A. Recio-Blanco\inst{17}
         \and
         M.~T. Costado\inst{18}
         \and
         A. Hourihane\inst{10}
         \and 
         C. Lardo\inst{19}
         \and
         P. de Laverny\inst{17}
         \and
         J. Lewis\inst{10}
         \and
         L. Monaco\inst{20}
         \and
         L. Morbidelli\inst{2}
         \and
         G.~G. Sacco\inst{2}
         \and
         C.~C. Worley\inst{10}
         \and
         S. Zaggia\inst{5}
         \and
         S. Martell\inst{21}
%          \fnmsep\thanks{Just to show the usage of the elements in the author field}
          }

\institute{
           Department for Astrophysics, Nicolaus Copernicus Astronomical Center, 
           ul. Rabia\'nska 8, 87-100 Toru\'n, Poland \\
           \email{rsmiljanic@camk.edu.pl}
           \and
           INAF - Osservatorio Astrofisico di Arcetri, Largo Enrico Fermi 5, 50125 Firenze, Italy
           \and
           INAF - Osservatorio Astronomico di Bologna, Via Ranzani 1, I-40127 Bologna, Italy
          \and
           European Southern Observatory, Karl-Schwarzschild-Str. 2, 85748 Garching bei M\"unchen, Germany
           \and
           INAF - Osservatorio Astronomico di Padova, Vicolo Osservatorio 2, I-35122 Padova, Italy
           \and
           Institute of Theoretical Physics and Astronomy, Vilnius University, Go{\v s}tauto 12, Vilnius LT-01108, Lithuania   
           \and
           INAF - Osservatorio Astrofisico di Catania, via S. Sofia 78, I-95123 Catania, Italy
           \and
           Dipartimento di Astronomia, Universit\`a di Bologna, Via Ranzani 1, I-40127 Bologna, Italy  
           \and
           Instituto de Astrof\'{\i}sica e Ci\^encias do Espa\c co, Universidade do Porto, Rua das Estrelas, 4150-762 Porto, Portugal
          \and
          Institute of Astronomy, University of Cambridge, Madingley Road, Cambridge CB3 0HA, United Kingdom
            \and
            Departamento de Astronom\'{\i}a, Universidad de Concepci\'on, Casilla 160-C, Concepci\'on, Chile
            \and
            Lund Observatory, Department of Astronomy and Theoretical Physics, Box 43, SE-221 00 Lund, Sweden
            \and
            GEPI, Observatoire de Paris, PSL Research University, CNRS, Univ Paris Diderot, Sorbonne Paris Cit\'e, 61 Avenue de l'Observatoire, 75014 Paris, France
           \and 
           Universit\'e de Picardie Jules Verne, Physics Dpt. 33 rue St Leu, F-80000 Amiens, France
            \and
            Dipartimento di Fisica e Astronomia, Sezione Astrofisica, Universit\'{a} di Catania, via S. Sofia 78, 95123, Catania, Italy
            \and
            ASI Science Data Center, Via del Politecnico SNC, 00133 Roma, Italy
            \and
            Laboratoire Lagrange, Universit\'e C\^ote d'Azur, Observatoire de la C\^ote d'Azur, CNRS, Bd de l'Observatoire, CS 34229, 06304 Nice cedex 4, France
            \and
            Instituto de Astrof\'{i}sica de Andaluc\'{i}a-CSIC, Apdo. 3004, 18080 Granada, Spain
            \and
            Astrophysics Research Institute, Liverpool John Moores University, 146 Brownlow Hill, Liverpool L3 5RF, United Kingdom
            \and
            Departamento de Ciencias Fisicas, Universidad Andres Bello, Republica 220, Santiago, Chile
            \and
            School of Physics, University of New South Wales, Sydney NSW 2052, Australia
           }

   \date{Received one day; accepted some time later}

\titlerunning{Inhibited extra mixing in two giants of Trumpler 20?}
\authorrunning{Smiljanic et al.}

% \abstract{}{}{}{}{} 
% 5 {} token are mandatory
 
  \abstract
 {}  
 {We report the discovery of two Li-rich giants, with A(Li) $\sim$ 1.50, in an analysis of a sample of 40 giants of the open cluster Trumpler 20 (with turnoff mass $\sim$ 1.8 $M_{\sun}$). The cluster was observed in the context of the Gaia-ESO Survey.}
 {The atmospheric parameters and Li abundances were derived using high-resolution UVES spectra. The Li abundances were corrected for nonlocal thermodynamical equilibrium (non-LTE) effects.}
 {Only upper limits of the Li abundance could be determined for the majority of the sample. Two giants with detected Li turned out to be Li rich: star MG 340 has A(Li)$_{\rm non-LTE}$ = 1.54 $\pm$ 0.21 dex and star MG 591 has A(Li)$_{\rm non-LTE}$ = 1.60 $\pm$ 0.21 dex. Star MG 340 is on average $\sim$ 0.30 dex more rich in Li than stars of similar temperature, while for star MG 591 this difference is on average $\sim$ 0.80 dex. Carbon and nitrogen abundances indicate that all stars in the sample have completed the first dredge-up.}
 {The Li abundances in this unique sample of 40 giants in one open cluster clearly show that extra mixing is the norm in this mass range. Giants with Li abundances in agreement with the predictions of standard models are the exception. To explain the two Li-rich giants, we suggest that all events of extra mixing have been inhibited. This includes rotation-induced mixing during the main sequence and the extra mixing at the red giant branch luminosity bump. Such inhibition has been suggested in the literature to occur because of fossil magnetic fields in red giants that are descendants of main-sequence Ap-type stars.}


   \maketitle
%
%________________________________________________________________

\section{Introduction}

Although not well understood, the phenomenon of Li-rich giants seems to be ubiquitous as they have been observed in different environments: open clusters, globular clusters, metal-rich and metal-poor field stars, the Galactic bulge, and also in dwarf galaxies \citep[see, e.g.,][and references therein]{1999A&A...348L..21H,2009A&A...508..289G,2011ApJ...730L..12K,2011ApJ...743..107R,2012ApJ...752L..16K,2016ApJ...819..135K,2015ApJ...801L..32D}.

Lithium-rich giants are usually defined as those that, after the first dredge-up, have A(Li) $\geq$ 1.50 dex. This limit is the post-dredge-up Li abundance of a low-mass star according to standard evolutionary models, i.e., those models that include only convection as a mixing mechanism. The first Li-rich giant was a fortuitous discovery by \citet{1982ApJ...255..577W}. Subsequent searches have shown that these stars comprise about 1--2\% of red giants \citep{1989ApJS...71..293B,2000AJ....119.2895P}. \citet{2000A&A...359..563C} suggested that these objects appear at the luminosity bump of the red giant branch (RGB) or at the early-asymptotic giant branch (AGB) for low- and intermediate-mass stars, respectively. Other recent results preferably classify these objects as core helium-burning giants \citep{2011ApJ...730L..12K,2014A&A...564L...6M,2014ApJ...784L..16S}. Nevertheless, some Li-rich giants have been found throughout the RGB \citep[see, e.g.,][]{2011A&A...531L..12A,2011A&A...529A..90M,2013MNRAS.430..611M}. 

\begin{figure}
\centering
\includegraphics[height = 8.5cm]{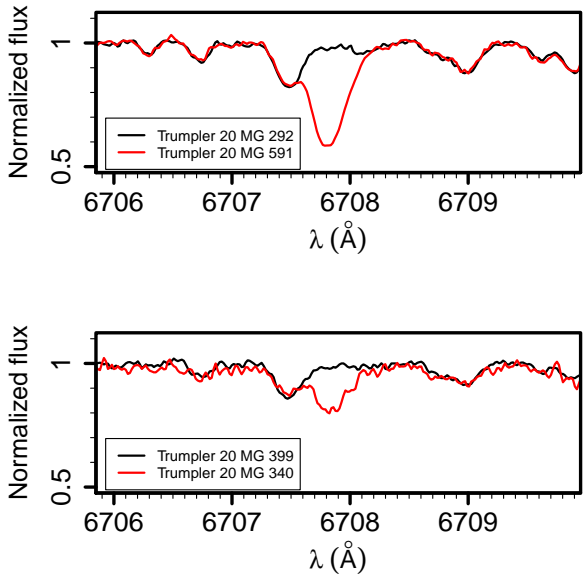}
\caption{Comparison of the spectra around the Li 6708 \AA\ line between the Li-rich giants and giants with similar atmospheric parameters.}\label{fig:spec}
\end{figure}

Lithium-rich giants have other noteworthy characteristics that add complexity to the puzzle. Some present a far-infrared excess, suggesting a connection with enhanced mass loss \citep{1996ApJ...456L.115D}. This mass-loss event can also explain the observation of complex organic and inorganic compounds detected in the infrared spectra of some Li-rich giants \citep{2015ApJ...806...86D}. In a few cases, the presence of circumstellar material has been confirmed by polarimetry \citep{2006A&A...449..211P}. Nevertheless, not all Li-rich giants have an infrared excess \citep[see, e.g.,][]{1999A&A...342..831J,2015A&A...577A..10B,2015AJ....150..123R}.

\citet{1993ApJ...403..708F} proposed a connection between Li enrichment, fast rotation, and chromospheric activity. Lithium-rich giants seem to be more common among fast rotating stars \citep[$\sim$ 50\%; see, e.g.,][]{2002AJ....123.2703D}. A strong magnetic field was detected in one Li-rich giant by \citet{2009A&A...504.1011L}. Further examples of Li-rich, fast-rotating, active giants exist \citep[e.g.,][]{2002AJ....123.1993R,2014A&A...571A..74K,2015A&A...574A..31S}.

In addition, a few Li-rich giants hosting planets have been found \citep[e.g.,][]{2012ApJ...754L..15A,2014A&A...569A..55A}. As proposed by \citet{1999MNRAS.308.1133S}, surface Li enrichment could be caused by planet engulfment, which also causes spin-up, magnetic field generation, and shell ejection. However, planet accretion would create a $^{9}$Be enhancement that has never been detected in Li-enriched objects \citep{1997A&A...321L..37D,1999A&A...345..249C,2005A&A...439..227M,2014A&A...563A...3P} with the exception of one F-type dwarf in the open cluster NGC 6633 \citep{2005MNRAS.363L..81A}. Alternatively, planet engulfment could activate internal Li production and induce its mixing to the surface \citep[][]{2000A&A...358L..49D}.

Indeed, the properties of many Li-rich giants discovered within the Gaia-ESO Survey \citep{2012Msngr.147...25G,2013Msngr.154...47R} seem to be consistent with those of giants that engulfed close-in giant planets before evolving up the RGB \citep{2016arXiv160303038C}. However, a small fraction of cases still require alternative explanations. Here, we report the discovery of two Li-rich giants that could be examples of such an alternative formation channel in the open cluster \object{Trumpler 20}, which is a system of $\sim$ 1.66 Gyr in age and [Fe/H] = +0.17 \citep[][]{2014A&A...561A..94D}. 

\begin{figure}
\centering
\includegraphics[height = 7cm]{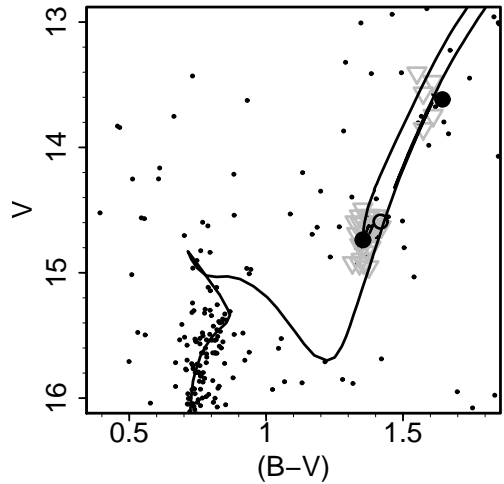}
\caption{Color magnitude diagram of Trumpler 20. Only stars within 3\arcmin of the cluster center are shown. The two Li-rich giants are shown as filled circles, other giants with Li detections as open circles, giants with Li upper limits as gray triangles, and the remaining stars in the field as dots. The solid line is an isochrone from \citet{2012MNRAS.427..127B} with age = 1.66 Gyr and [Fe/H] = +0.17, which is the best fit to the photometric data as determined by \citet{2014A&A...561A..94D}.}\label{fig:cmd}
\end{figure}

This paper is organized as follows. In Sect.\ \ref{sec:data} we briefly describe the data used here, the analysis, and the properties of our sample. In Sect.\ \ref{sec:extra} we discuss how extra mixing is needed to explain the surface Li abundances of the majority of the sample. In Sect.\ \ref{sec:noextra} we discuss the two Li-rich giants and the possibility that they have avoided extra-mixing mechanisms. Finally, Sect.\ \ref{sec:end} summarizes our findings and suggests new observations that could support our interpretation of the Li enrichment in these two giants.

%________________________________________________________________
\section{Data, analysis, and sample properties}\label{sec:data}

The high-resolution (R $\sim$ 47\,000)  UVES \citep[Ultraviolet and Visual Echelle Spectrograph,][]{2000SPIE.4008..534D} spectra of 42 targets in Trumpler 20 were obtained in the context of the Gaia-ESO Survey. Data reduction is described in \citet{2014A&A...565A.113S}. Basic information on the observed giants is available online in Table \ref{tab:obs}. 

The atmospheric parameters and abundances (see online Table \ref{tab:par}) are part of the fourth Gaia-ESO internal data release (iDR4). The spectra were analyzed using the Gaia-ESO multiple pipelines strategy \citep{2014A&A...570A.122S} with an updated methodology (Casey et al. 2016b, in preparation).

Membership was assigned using the radial velocities (RVs) as in \citet{2014A&A...561A..94D}. Likely members (40 giants in total) are those with RV within three standard deviations of the cluster average ($\overline{RV}$ $\pm \ \sigma$ = $-$40.2 $\pm$ 1.3 km s$^{-1}$). One star is a subgiant\footnote{Trumpler 20 MG 430. The numbering system we adopt is that defined in \citet{2005ApJS..161..118M}.} and one a probable nonmember (or binary) with deviant RV\footnote{Trumpler 20 MG 894 with RV = $-$35.3 km s$^{-1}$.}.
 
The Li abundances were determined from the 6708 \AA\ line. In Fig.\ \ref{fig:spec}, we compare the Li 6708 \AA\ lines of the two Li-rich giants to those of stars with similar atmospheric parameters to illustrate the Li enhancement. Corrections for nonlocal thermodynamical equilibrium (non-LTE) effects were applied using the grid of \citet{2009A&A...503..541L}. For the giants, the corrections range from 0.14 dex to 0.32 dex, depending on the atmospheric parameters.

The color magnitude diagram (CMD) of Trumpler 20 is shown in Fig.\ \ref{fig:cmd}. The photometry is originally from \citet{2010AJ....140..954C} corrected for differential reddening by \citet{2014A&A...561A..94D}. The uncertainties in the magnitudes are $\sim$0.02-0.05 mag. A noticeable feature in this CMD is the extended clump region of the cluster. Trumpler 20 is well known for its peculiar extended clump region \citep[see][]{2010AJ....140..954C,2012ApJ...751L...8P,2014A&A...561A..94D}. This feature is probably caused by the presence of two distinct clumps; the fainter clump comprises stars massive enough to start core He-burning in nondegenerate conditions and the brighter clump comprises stars with slightly lower mass that have been through the He-core flash \citep[see, e.g.,][]{1999MNRAS.308..818G,2000A&A...354..892G}.

\begin{figure}
\centering
\includegraphics[height = 7cm]{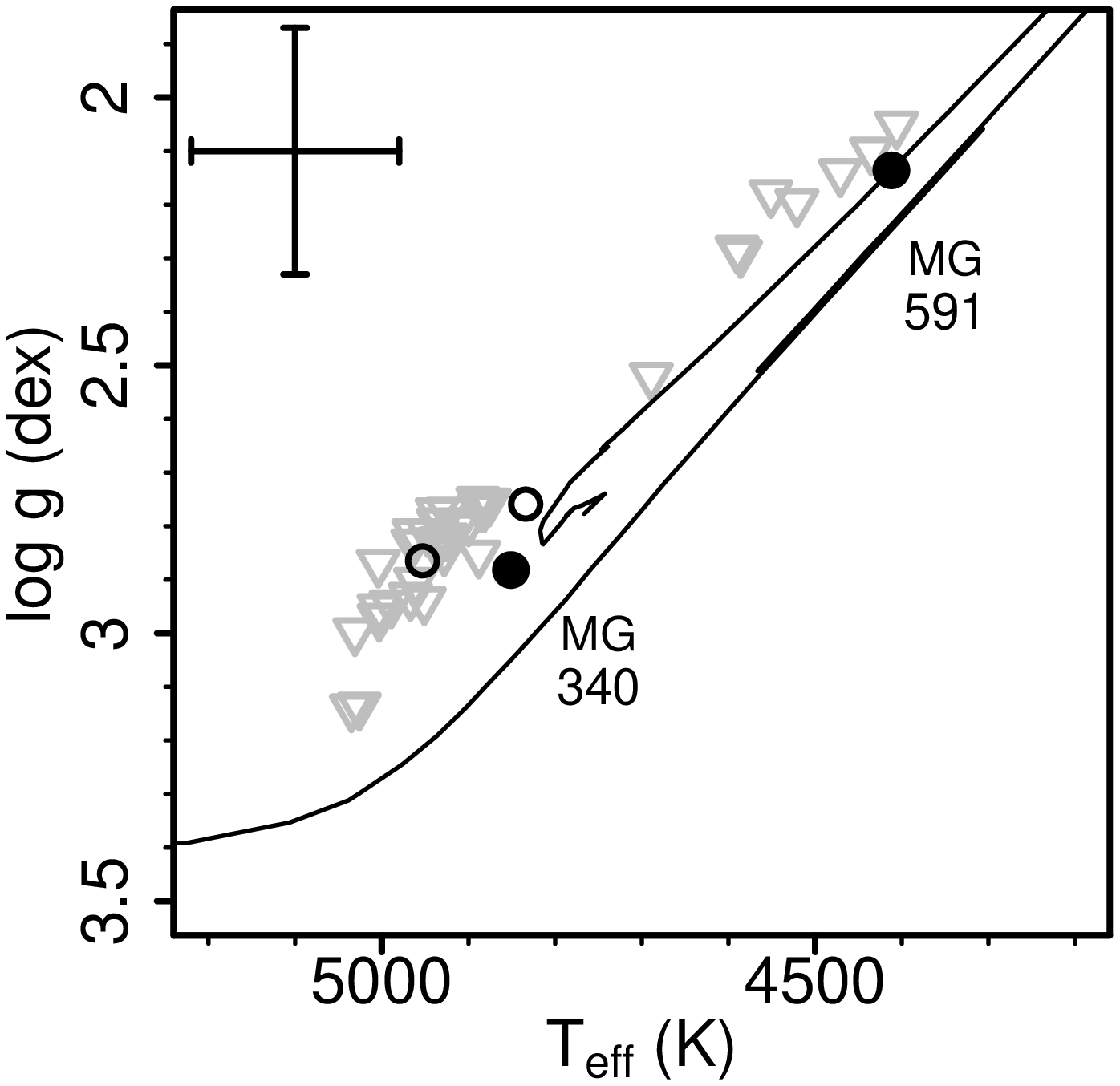}
\caption{Trumpler 20 giants in the $T_{\rm eff}$-$\log g$ plane. The two Li-rich giants are shown as filled circles, other giants with Li detections as open circles, and giants with Li upper limits as gray triangles. The solid line is an isochrone from \citet{2012MNRAS.427..127B} of 1.66 Gyr in age and [Fe/H] = +0.17, which is the best fit to the photometric data \citep[by][]{2014A&A...561A..94D}. A typical error bar ($\pm$ 120 K and $\pm$ 0.23 dex for $T_{\rm eff}$ and $\log g$, respectively) is shown in the top left.}\label{fig:tefflogg}
\end{figure}

Figure \ref{fig:tefflogg} shows the sample in the $T_{\rm eff}$-$\log~g$ plane. The group of giants with lower $\log~g$ are either at the luminosity bump of the RGB or at the early-AGB, as both stages are within the error bars of the parameters in Fig.\ \ref{fig:tefflogg} and are hard to separate in the CMD of Fig.\ \ref{fig:cmd}. We can be more confident about the evolutionary state of the group of giants with higher $\log~g$ because of their chemical abundances.

\begin{figure}
\centering
\includegraphics[height = 7cm]{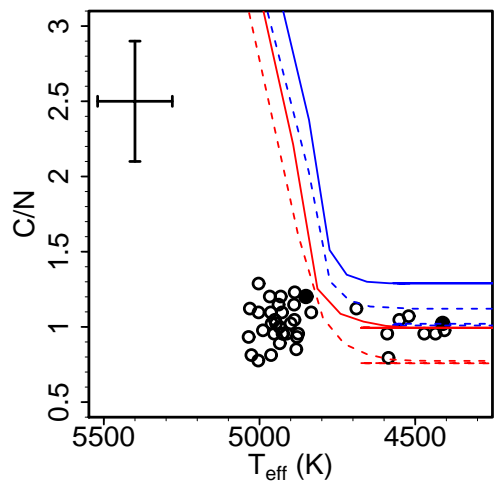}
\caption{C/N ratio as a function of T$_{\rm eff}$. Solid lines are the predictions of standard models and dashed lines of models with rotation-induced mixing and thermohaline mixing \citep{2012A&A...543A.108L}. Lines in blue and red are for solar metallicity stars of 1.5 M$_{\odot}$ and 2.0 M$_{\odot}$, respectively. The two Li-rich giants are shown as full circles.}\label{fig:cnteff}
\end{figure}

The evolutionary stage of the stars is an important source of information to interpret their Li abundances, as a high Li abundance could just be indicating that the giant is actually at the bottom of the RGB \citep[e.g.,][]{2015MNRAS.446.3562B}. Nevertheless, the C and N abundances of the giants demonstrate that they have all completed the standard first dredge-up. As shown in Fig. \ref{fig:cnteff}, according to the models of \citet{2012A&A...543A.108L}, giants of 1.5 and 2.0 M$_{\odot}$ after the dredge-up have C/N $\sim$ 1, as do all the giants in our sample. Giants at the bottom of the RGB with $T_{\rm eff}$ $\sim$ 5000 K would be in the stage before the end of the first dredge-up and thus would instead have C/N > 3.  We can thus safely conclude that i) all the giants with $T_{\rm eff}$ $\sim$ 5000 K are in the red clump and not on the RGB and ii) that all the brightest and coolest giants have completed the Li dilution expected during the first dredge-up. The C and N abundances of the Trumpler 20 giants were discussed in \citep[][]{2015A&A...573A..55T}. 

%

%________________________________________________________________

\section{Extra mixing in the majority of the giants}\label{sec:extra}

Lithium abundances have been extensively used as a tracer of mixing processes, as Li is rapidly destroyed in (p,$\alpha$) reactions at temperatures above 2.5 $\times 10^{6}$ K \citep{1951ApJ...113..536G}. Thus, Li only survives in the outermost layers of a star. As stars evolve to the RGB, the convective envelope deepens and Li from the surface is diluted.

\begin{figure}
\centering
\includegraphics[height = 7cm]{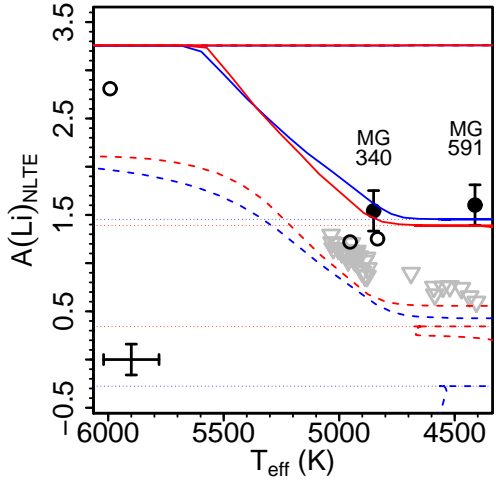}
\caption{Lithium abundance as a function of T$_{\rm eff}$. The two Li-rich giants are shown as filled circles, other giants with Li detections as open circles, and giants with Li upper limits as gray triangles. Solid lines are the predictions of standard models and dashed lines of models with rotation-induced mixing and thermohaline mixing \citep{2012A&A...543A.108L}. Lines in blue and red are for solar metallicity stars of 1.5 M$_{\odot}$ and 2.0 M$_{\odot}$, respectively. The dotted lines are included as an eye guide to the Li abundance level of clump giants in the models (but we note here that the clump phase does not span the T$_{\rm eff}$ range of the dotted lines in the plot).}\label{fig:liteff}
\end{figure}

Figure \ref{fig:liteff} shows the Li abundances as a function of the effective temperatures (T$_{\rm eff}$), for all cluster members, in comparison with the models of \citet{2012A&A...543A.108L}. The bulk of the stars fall in between the standard and extra-mixing models (solid and dashed lines, respectively). However, care is needed in interpreting the plot because of the evolutionary state of the stars. 

The solid and dashed lines in Fig.\ \ref{fig:liteff} are predictions for first ascent RGB stars and not for clump giants. As we showed above, our giants with $T_{\rm eff}$ $\sim$ 5000 K are clump giants and not first ascent RGB stars. The observations should not be compared to this region of the models, but rather to the A(Li) level of core-He burning giants (the dotted lines). The clump giants have Li upper limits on average of about 0.3 dex below the two top dotted lines (standard models). For lower temperatures, the second group of stars has also Li upper limits well below the prediction of the standard models. 

The clear exceptions to that are the two Li-rich giants, Trumpler 20 MG 340 and 591. This agrees with the findings of \citet{1989ApJS...71..293B} that giants with A(Li) $\sim$ 1.50 in agreement with standard models are a minority. Here, we are able to confirm this result in a large sample of giants of the same age, same initial chemical composition, and very similar masses. The enhanced Li depletion/dilution seen in the majority of the giants of Trumpler 20 is well documented in the literature \citep[e.g.,][]{1999A&A...345..936L,2001A&A...374.1017P,2004A&A...424..951P}, although the mechanism behind this extra mixing is still under debate.

We note another possible outlier, star MG 505, with A(Li)$_{\rm non-LTE}$ = 1.25 $\pm$ 0.21. However, because within the errors its abundance agrees with the highest upper limit at its temperature, we do not consider it among those that agree with the standard models. 

%
%________________________________________________________________

\section{Inhibited extra mixing in two giants}\label{sec:noextra}

Star MG 340 has A(Li)$_{\rm non-LTE}$ = 1.54 $\pm$ 0.21 and T$_{\rm eff}$ = 4851 K, while in eight other stars with T$_{\rm eff}$ = 4850 $\pm$ 50 K there is one detection at A(Li)$_{\rm non-LTE}$ = 1.25 and seven upper limits below  A(Li)$_{\rm non-LTE} \sim$ 1.05. Star MG 591 has A(Li)$_{\rm non-LTE}$ = 1.60 $\pm$ 0.21 and T$_{\rm eff}$ = 4412 K, while seven other stars with T$_{\rm eff}$ < 4600 K have upper limits below A(Li)$_{\rm non-LTE} \sim$ 0.76. 

These two Li-rich giants do not show any additional chemical peculiarity when compared to the other cluster giants. Both stars seem to be single, but we do not have multiple epoch spectra to exclude long period companions. All sample giants seem to be slow rotators: $v~\sin~i < 4.0$  km s$^{-1}$. Therefore, the Li enhancement is probably not connected to rotation in the sense seen by \citet{2006A&A...450.1173L} and \citet{2012ApJ...757..109C}. These authors found that in a given sample of giants, those with higher Li abundance tend to be those giants that are rotating faster, however, our giants might seem to be slow rotators because of an unfavorable line of sight. Slow rotation also argues against, but does not fully exclude, planet accretion with transfer of angular momentum as the source of the Li enhancement \citep[see also][]{2016ApJ...818...25C,2016A&A...587A..66D}. Thus, external pollution as advocated by \citet{2016arXiv160303038C} to explain other Li-rich giants discovered within the Gaia-ESO Survey seems unlikely in our case.

Internal Li production was also suggested to explain Li-rich giants. Fresh Li production might occur in the stellar interior through the $^{7}$Be mechanism \citep{1971ApJ...164..111C}. However, it is still unknown which transport mechanism would bring $^{7}$Be, which decays to $^{7}$Li, to the surface \citep{1999ApJ...510..217S,2001A&A...375L...9P}. \citet{2000A&A...359..563C} argued that Li-rich giants were preferentially found at the bump and the early-AGB, and connected the Li enrichment with an extra-mixing process that activates at these evolutionary stages. For low-mass stars at the bump, the extra mixing would be connected to the H-burning shell that is moving outward and removes the molecular weight discontinuity left by the receding convective layer. In intermediate-mass stars, the extra mixing would take place at the early-AGB when the convective envelope deepens again.

The two Li-rich clump giants found by \citet{2014ApJ...784L..16S}, with A(Li) = 2.71, and \citet{2014A&A...564L...6M}, with A(Li) = 3.75, showed that the above scenario is at least incomplete. For clump giants, the Li enrichment could be connected to the He flash, following an episode of H injection in deeper high temperature regions \citep{2011ApJ...743...55M}.

Because of the large surface convective layers of giants, the observed Li surface enrichment is likely to be short lived. A Li-rich low-mass giant that appears at the bump should not remain Li rich during its evolution to the clump. More likely, the two Li-rich giants discovered here have been freshly created during or close to their current evolutionary stages. Thus, to explain our Li-rich giants with internal Li production would require a combination of the two distinct scenarios above. One Li-rich giant would be created by mixing induced by the He flash, the other would be created by extra mixing at the bump. It also seems an odd coincidence that we would happen to observe both Li-rich giants at the moment in which their Li abundances are very close to the value expected by standard models.

While internal Li production could indeed be required to explain abundances above the meteoritic value (A(Li) $>$ 3.0), this is perhaps not necessary to explain stars MG 340 and 591. We instead suggest that their higher Li abundance is the result of the inhibition of extra-mixing processes. Without extra mixing, their surface Li abundance is at the level predicted by standard models. In addition, this single scenario would be able to explain both giants regardless of their different evolutionary states.

If this suggestion is correct, two instances of extra mixing must have been affected. The first is the extra Li dilution beyond the predicted first dredge-up dilution, which is observed as the star leaves the main sequence toward the RGB. Observationally, it is well known that an extra mixing causes the Li dilution to start earlier than predicted by standard models \citep[e.g.,][]{1999A&A...345..936L,2001A&A...374.1017P,2004A&A...424..951P}. This is because rotation-induced mixing creates a Li-free region inside real stars that is larger than predicted by these models \citep[]{2003A&A...399..603P}. In the two Li-rich giants, rotation-induced mixing must have been weak and the Li dilution proceeded as expected by standard models.

The second extra-mixing event to be avoided is the event taking place at the luminosity bump of the RGB \citep[see][and references therein]{2012A&A...543A.108L}. The luminosity bump \citep[see, e.g.,][]{2015MNRAS.453..666C} happens at the RGB of low-mass stars when the hydrogen-burning shell reaches the composition discontinuity left behind by the first dredge-up. The current best candidate for the mechanism behind this extra mixing seems to be thermohaline mixing \citep{2007A&A...467L..15C}, although there are discussions about the physical properties and efficiency of this mechanism \citep[e.g.,][and references therein]{2013A&A...553A...1M,2015ApJ...815...42G,2015MNRAS.446.2673L}.

Star MG 340 in the extended clump of the cluster is likely to be a low-mass star. As discussed by \citet{1999MNRAS.308..818G}, in such an extended clump, the less massive stars are actually the brighter stars. Both the CMD in Fig.\ \ref{fig:cmd} and the $T_{\rm eff}$-$\log~g$ diagram in Fig. \ref{fig:tefflogg} seem to indicate that MG 340 belongs to the group of brighter giants. Thus, in our scenario, for it to keep an unaltered Li surface abundance, thermohaline mixing must have been inhibited. On the other hand, star MG 591 is either on the bump or on the early-AGB. If on the early-AGB, then it is an intermediate-mass star that does not go through both the He-flash and the bump phase. However, it would still need to inhibit thermohaline mixing at the early-AGB \citep[see][and references therein]{2012A&A...543A.108L}. If on the bump, it either avoided thermohaline mixing or did not activate it yet. All other giants with similar $T_{\rm eff}$ and $\log g$ have lower Li abundances. This could indicate that star MG 591 is also after the moment where thermohaline mixing becomes efficient. As pointed out by the referee, however, at this phase stars ascend and reascend the RGB, crossing the same $T_{\rm eff}$ and $\log g$ region three times. Therefore, it is plausible that at least one star among the group at the bump has not yet activated thermohaline mixing. If this is the case, star MG 591 would not be an Li-rich giant, but a normal giant in a stage before extra mixing was activated.

Extra-mixing inhibition is not a new idea. Based on carbon isotopic ratios, $^{12}$C/$^{13}$C, \citet{1998A&A...336..915C} estimated that about 4\% of low-mass giants do not experience extra mixing on the RGB. \citet{2007A&A...476L..29C} suggested that extra mixing is avoided by giants that are descendant from Ap-type main-sequence stars. In these stars, fossil magnetic fields would be able to inhibit thermohaline mixing. Modern estimates of the percentage of Ap stars with respect to nonmagnetic A-type stars are between 1.7-3.5\% \citep{1993ASPC...44..577N,2007pms..conf...89P}. This is consistent with finding one or two stars in our sample of 40 giants of Trumpler 20 (a fraction of 2.5 or 5\%). In addition, the stellar mass of our giants is within the mass range of Ap stars \citep[$\sim$ 1.5 to 3.6 M$_{\odot}$,][]{2007pms..conf...89P}.

%
%________________________________________________________________

\section{Summary}\label{sec:end}

In this work, we presented the discovery of two Li-rich giants in the open cluster Trumpler 20. These two stars were identified in an analysis of a sample of 40 giants for which high-resolution spectra were obtained with UVES in the context of the Gaia-ESO public spectroscopic survey. This provides a unique large sample of giants that have the same age, same initial chemical composition, and very similar masses. The Li abundances in this sample clearly demonstrate that extra mixing is the norm in stars in this mass range. Giants with Li abundances in agreement with the predictions of standard models are the exception. 

To explain the two Li-rich giants, we suggest that all instances of extra-mixing processes have been inhibited. Because of that, the surface Li abundance in these two stars remained at the level predicted by standard stellar evolution models, i.e., A(Li) $\sim$ 1.50. We argue that the fraction of Li-rich giants found in our sample is consistent with these giants being evolved counterparts of magnetic Ap-type dwarfs. In this case, the extra-mixing processes could have been inhibited by the action of magnetic fields, as suggested by \citet{2007A&A...476L..29C}. 

Other explanations seem less likely, although they cannot be fully excluded. Extra Li from the accretion of external material should be accompanied by accretion of angular momentum, but there is no evidence of fast rotation in the two giants. Because the two Li-rich giants have different evolutionary stages, internal Li production would require two different mechanisms to bring the fresh Li to the surface. The extra-mixing inhibition hypothesis would instead be able to explain both giants at the same time.

Additional observations could help in providing extra support to our suggested scenario, or they could help to disprove it. First, if no extra mixing took place, the carbon isotopic ratio should be close to the prediction of standard models, $^{12}$C/$^{13}$C $\sim$ 30. We could not determine the carbon isotopic ratio because the region around 8000\AA, containing the CN bands preferred for this type of analysis in metal-rich giants, is not part of the Gaia-ESO spectra. 

Second, there should likely be signs of magnetic activity in the Li-rich giants. Other candidates of Ap-type stars descendants were identified among giants and subgiants with magnetic activity \citep[e.g.,][]{2014psce.conf..444A}. However, fossil magnetic fields beneath the surface are hard to detect \citep{2015A&A...574A..90A,2016Natur.529..364S}. Nevertheless, the core-He burning star MG 340 is at one of the evolutionary phases where magnetic activity in giants is observed \citep{2015A&A...574A..90A}. We checked the H$\alpha$ line in our spectrum, but it shows no evident sign of activity. This is not inconsistent, as not all active giants display emission in H$\alpha$ \citep[see, e.g.,][]{1993ApJ...403..708F}. Emission should be clearer in the Ca H and K lines or in the UV, which are not part of our spectra.

Finally, if such a scenario of extra-mixing inhibition is correct, it would likely apply to many, if not all, Li-rich giants with A(Li) $\sim$ 1.50 (or slightly higher) and masses between $\sim$ 1.5 and 3.6 M$_{\odot}$. They would not have experienced fresh Li production, but would instead have preserved part of their original Li abundance.

\begin{acknowledgements}
We thank the anonymous referee for his/her suggestions and very fast report. R.S. acknowledges support by the National Science Center of Poland through grant 2012/07/B/ST9/04428. S.V. gratefully acknowledges the support provided by FONDECYT reg. n. 1130721. G.T. acknowledges support by the Research Council of Lithuania (MIP-082/2015). D.G. gratefully acknowledges support from the Chilean BASAL Centro de Excelencia en Astrof\'{\i}sica y Tecnolog\'{\i}as Afines (CATA) grant PFB-06/2007. S.G.S. acknowledges the support from FCT through Investigador FCT contract of reference IF/00028/2014. E.D.M. acknowledges the support from FCT in the form of the grant SFRH/BPD/76606/2011. S.G.S. and E.D.M. also acknowledge the support from FCT through the project PTDC/FIS-AST/7073/2014. Based on data products from observations made with ESO Telescopes at the La Silla Paranal Observatory under program ID 188.B-3002. These data products have been processed by the Cambridge Astronomy Survey Unit (CASU) at the Institute of Astronomy, University of Cambridge, and by the FLAMES/UVES reduction team at INAF/Osservatorio Astrofisico di Arcetri. These data have been obtained from the Gaia-ESO Survey Data Archive, prepared and hosted by the Wide Field Astronomy Unit, Institute for Astronomy, University of Edinburgh, which is funded by the UK Science and Technology Facilities Council. This work was partly supported by the European Union FP7 program through ERC grant number 320360 and by the Leverhulme Trust through grant RPG-2012-541. We acknowledge the support from INAF and Ministero dell' Istruzione, dell' Universit\`a' e della Ricerca (MIUR) in the form of the grant ``Premiale VLT 2012'' and ``The Chemical and Dynamical Evolution of the Milky Way and Local Group Galaxies''. The results presented here benefit from discussions held during the Gaia-ESO workshops and conferences supported by the ESF (European Science Foundation) through the GREAT Research Network Programme.
\end{acknowledgements}

\bibliographystyle{aa} % style aa.bst
\bibliography{../../smiljanic} % your references Yourfile.bib

\begin{thebibliography}{80}
\expandafter\ifx\csname natexlab\endcsname\relax\def\natexlab#1{#1}\fi

\bibitem[{{Adam{\'o}w} {et~al.}(2012){Adam{\'o}w}, {Niedzielski}, {Villaver},
  {Nowak}, \& {Wolszczan}}]{2012ApJ...754L..15A}
{Adam{\'o}w}, M., {Niedzielski}, A., {Villaver}, E., {Nowak}, G., \&
  {Wolszczan}, A. 2012, \apjl, 754, L15

\bibitem[{{Adam{\'o}w} {et~al.}(2014){Adam{\'o}w}, {Niedzielski}, {Villaver},
  {Wolszczan}, \& {Nowak}}]{2014A&A...569A..55A}
{Adam{\'o}w}, M., {Niedzielski}, A., {Villaver}, E., {Wolszczan}, A., \&
  {Nowak}, G. 2014, \aap, 569, A55

\bibitem[{{Alcal{\'a}} {et~al.}(2011){Alcal{\'a}}, {Biazzo}, {Covino},
  {Frasca}, \& {Bedin}}]{2011A&A...531L..12A}
{Alcal{\'a}}, J.~M., {Biazzo}, K., {Covino}, E., {Frasca}, A., \& {Bedin},
  L.~R. 2011, \aap, 531, L12

\bibitem[{{Ashwell} {et~al.}(2005){Ashwell}, {Jeffries}, {Smalley},
  {Deliyannis}, {Steinhauer}, \& {King}}]{2005MNRAS.363L..81A}
{Ashwell}, J.~F., {Jeffries}, R.~D., {Smalley}, B., {et~al.} 2005, \mnras, 363,
  L81

\bibitem[{{Auri{\`e}re} {et~al.}(2015){Auri{\`e}re}, {Konstantinova-Antova},
  {Charbonnel}, {Wade}, {Tsvetkova}, {Petit}, {Dintrans}, {Drake}, {Decressin},
  {Lagarde}, {Donati}, {Roudier}, {Ligni{\`e}res}, {Schr{\"o}der},
  {Landstreet}, {L{\`e}bre}, {Weiss}, \& {Zahn}}]{2015A&A...574A..90A}
{Auri{\`e}re}, M., {Konstantinova-Antova}, R., {Charbonnel}, C., {et~al.} 2015,
  \aap, 574, A90

\bibitem[{{Auri{\`e}re} {et~al.}(2014){Auri{\`e}re}, {Ligni{\`e}res},
  {Konstantinova-Antova}, {Charbonnel}, {Petit}, {Tsvetkova}, \&
  {Wade}}]{2014psce.conf..444A}
{Auri{\`e}re}, M., {Ligni{\`e}res}, F., {Konstantinova-Antova}, R., {et~al.}
  2014, in Putting A Stars into Context: Evolution, Environment, and Related
  Stars, ed. G.~{Mathys}, E.~R. {Griffin}, O.~{Kochukhov}, R.~{Monier}, \&
  G.~M. {Wahlgren}, 444--450

\bibitem[{{Bharat Kumar} {et~al.}(2015){Bharat Kumar}, {Reddy},
  {Muthumariappan}, \& {Zhao}}]{2015A&A...577A..10B}
{Bharat Kumar}, Y., {Reddy}, B.~E., {Muthumariappan}, C., \& {Zhao}, G. 2015,
  \aap, 577, A10

\bibitem[{{B{\"o}cek Topcu} {et~al.}(2015){B{\"o}cek Topcu}, {Af{\c s}ar},
  {Schaeuble}, \& {Sneden}}]{2015MNRAS.446.3562B}
{B{\"o}cek Topcu}, G., {Af{\c s}ar}, M., {Schaeuble}, M., \& {Sneden}, C. 2015,
  \mnras, 446, 3562

\bibitem[{{Bressan} {et~al.}(2012){Bressan}, {Marigo}, {Girardi}, {Salasnich},
  {Dal Cero}, {Rubele}, \& {Nanni}}]{2012MNRAS.427..127B}
{Bressan}, A., {Marigo}, P., {Girardi}, L., {et~al.} 2012, \mnras, 427, 127

\bibitem[{{Brown} {et~al.}(1989){Brown}, {Sneden}, {Lambert}, \&
  {Dutchover}}]{1989ApJS...71..293B}
{Brown}, J.~A., {Sneden}, C., {Lambert}, D.~L., \& {Dutchover}, Jr., E. 1989,
  \apjs, 71, 293

\bibitem[{{Cameron} \& {Fowler}(1971)}]{1971ApJ...164..111C}
{Cameron}, A.~G.~W. \& {Fowler}, W.~A. 1971, \apj, 164, 111

\bibitem[{{Carlberg} {et~al.}(2012){Carlberg}, {Cunha}, {Smith}, \&
  {Majewski}}]{2012ApJ...757..109C}
{Carlberg}, J.~K., {Cunha}, K., {Smith}, V.~V., \& {Majewski}, S.~R. 2012,
  \apj, 757, 109

\bibitem[{{Carlberg} {et~al.}(2016){Carlberg}, {Smith}, {Cunha}, \&
  {Carpenter}}]{2016ApJ...818...25C}
{Carlberg}, J.~K., {Smith}, V.~V., {Cunha}, K., \& {Carpenter}, K.~G. 2016,
  \apj, 818, 25

\bibitem[{{Carraro} {et~al.}(2010){Carraro}, {Costa}, \&
  {Ahumada}}]{2010AJ....140..954C}
{Carraro}, G., {Costa}, E., \& {Ahumada}, J.~A. 2010, \aj, 140, 954

\bibitem[{{Casey} {et~al.}(2016){Casey}, {Ruchti}, {Masseron}, {Randich},
  {Gilmore}, {Lind}, {Kennedy}, {Smiljanic}, {Jofre}, {Hourihane}, {Bragaglia},
  {Donati}, {Koposov}, {Worley}, {Lanzafame}, {Franciosini}, {Lewis},
  {Magrini}, {Morbidelli}, {Sacco}, {Bensby}, {Carraro}, {Damiani}, {Feltzing},
  {Jeffries}, {Korn}, {Lardo}, {de Laverny}, {Delgado Mena}, {Costado},
  {Flaccomio}, {Frasca}, {Monaco}, {Pancino}, {Prisinzano}, {Recio-Blanco},
  {Sbordone}, {Sousa}, {Tautvaisiene}, {Vallenari}, {Zaggia}, {Zwitter}, {Silva
  Aguirre}, {Chorniy}, {Martell}, {Miglo}, {Chiappini}, {Montalban}, {Morel},
  \& {Valentini}}]{2016arXiv160303038C}
{Casey}, A.~R., {Ruchti}, G., {Masseron}, T., {et~al.} 2016, ArXiv e-prints,
  1603.03038

\bibitem[{{Castilho} {et~al.}(1999){Castilho}, {Spite}, {Barbuy}, {Spite}, {de
  Medeiros}, \& {Gregorio-Hetem}}]{1999A&A...345..249C}
{Castilho}, B.~V., {Spite}, F., {Barbuy}, B., {et~al.} 1999, \aap, 345, 249

\bibitem[{{Charbonnel} \& {Balachandran}(2000)}]{2000A&A...359..563C}
{Charbonnel}, C. \& {Balachandran}, S.~C. 2000, \aap, 359, 563

\bibitem[{{Charbonnel} \& {Do Nascimento}(1998)}]{1998A&A...336..915C}
{Charbonnel}, C. \& {Do Nascimento}, Jr., J.~D. 1998, \aap, 336, 915

\bibitem[{{Charbonnel} \& {Zahn}(2007{\natexlab{a}})}]{2007A&A...476L..29C}
{Charbonnel}, C. \& {Zahn}, J.-P. 2007{\natexlab{a}}, \aap, 476, L29

\bibitem[{{Charbonnel} \& {Zahn}(2007{\natexlab{b}})}]{2007A&A...467L..15C}
{Charbonnel}, C. \& {Zahn}, J.-P. 2007{\natexlab{b}}, \aap, 467, L15

\bibitem[{{Christensen-Dalsgaard}(2015)}]{2015MNRAS.453..666C}
{Christensen-Dalsgaard}, J. 2015, \mnras, 453, 666

\bibitem[{{de la Reza} {et~al.}(1996){de la Reza}, {Drake}, \& {da
  Silva}}]{1996ApJ...456L.115D}
{de la Reza}, R., {Drake}, N.~A., \& {da Silva}, L. 1996, \apjl, 456, L115

\bibitem[{{de la Reza} {et~al.}(2015){de la Reza}, {Drake}, {Oliveira}, \&
  {Rengaswamy}}]{2015ApJ...806...86D}
{de la Reza}, R., {Drake}, N.~A., {Oliveira}, I., \& {Rengaswamy}, S. 2015,
  \apj, 806, 86

\bibitem[{{de Medeiros} {et~al.}(1997){de Medeiros}, {Lebre}, {de Garcia Maia},
  \& {Monier}}]{1997A&A...321L..37D}
{de Medeiros}, J.~R., {Lebre}, A., {de Garcia Maia}, M.~R., \& {Monier}, R.
  1997, \aap, 321, L37

\bibitem[{{Dekker} {et~al.}(2000){Dekker}, {D'Odorico}, {Kaufer}, {Delabre}, \&
  {Kotzlowski}}]{2000SPIE.4008..534D}
{Dekker}, H., {D'Odorico}, S., {Kaufer}, A., {Delabre}, B., \& {Kotzlowski}, H.
  2000, in Society of Photo-Optical Instrumentation Engineers (SPIE) Conference
  Series, Vol. 4008, Optical and IR Telescope Instrumentation and Detectors,
  ed. M.~{Iye} \& A.~F. {Moorwood}, 534--545

\bibitem[{{Delgado Mena} {et~al.}(2016){Delgado Mena}, {Tsantaki}, {Sousa},
  {Kunitomo}, {Adibekyan}, {Zaworska}, {Santos}, {Israelian}, \&
  {Lovis}}]{2016A&A...587A..66D}
{Delgado Mena}, E., {Tsantaki}, M., {Sousa}, S.~G., {et~al.} 2016, \aap, 587,
  A66

\bibitem[{{Denissenkov} \& {Weiss}(2000)}]{2000A&A...358L..49D}
{Denissenkov}, P.~A. \& {Weiss}, A. 2000, \aap, 358, L49

\bibitem[{{Donati} {et~al.}(2014){Donati}, {Cantat Gaudin}, {Bragaglia},
  {Friel}, {Magrini}, {Smiljanic}, {Vallenari}, {Tosi}, {Sordo}, {Tautvai{\v
  s}ien{\.e}}, {Blanco-Cuaresma}, {Costado}, {Geisler}, {Klutsch}, {Mowlavi},
  {Mu{\~n}oz}, {San Roman}, {Zaggia}, {Gilmore}, {Randich}, {Bensby},
  {Flaccomio}, {Koposov}, {Korn}, {Pancino}, {Recio-Blanco}, {Franciosini}, {de
  Laverny}, {Lewis}, {Morbidelli}, {Prisinzano}, {Sacco}, {Worley},
  {Hourihane}, {Jofr{\'e}}, {Lardo}, \& {Maiorca}}]{2014A&A...561A..94D}
{Donati}, P., {Cantat Gaudin}, T., {Bragaglia}, A., {et~al.} 2014, \aap, 561,
  A94

\bibitem[{{D'Orazi} {et~al.}(2015){D'Orazi}, {Gratton}, {Angelou}, {Bragaglia},
  {Carretta}, {Lattanzio}, {Lucatello}, {Momany}, \&
  {Sollima}}]{2015ApJ...801L..32D}
{D'Orazi}, V., {Gratton}, R.~G., {Angelou}, G.~C., {et~al.} 2015, \apjl, 801,
  L32

\bibitem[{{Drake} {et~al.}(2002){Drake}, {de la Reza}, {da Silva}, \&
  {Lambert}}]{2002AJ....123.2703D}
{Drake}, N.~A., {de la Reza}, R., {da Silva}, L., \& {Lambert}, D.~L. 2002,
  \aj, 123, 2703

\bibitem[{{Fekel} \& {Balachandran}(1993)}]{1993ApJ...403..708F}
{Fekel}, F.~C. \& {Balachandran}, S. 1993, \apj, 403, 708

\bibitem[{{Garaud} \& {Brummell}(2015)}]{2015ApJ...815...42G}
{Garaud}, P. \& {Brummell}, N. 2015, \apj, 815, 42

\bibitem[{{Gilmore} {et~al.}(2012){Gilmore}, {Randich}, {Asplund}, {Binney},
  {Bonifacio}, {Drew}, {Feltzing}, {Ferguson}, {Jeffries}, {Micela},
  {Negueruela}, {Prusti}, {Rix}, {Vallenari}, {Alfaro}, {Allende-Prieto},
  {Babusiaux}, {Bensby}, {Blomme}, {Bragaglia}, {Flaccomio}, {Fran{\c c}ois},
  {Irwin}, {Koposov}, {Korn}, {Lanzafame}, {Pancino}, {Paunzen},
  {Recio-Blanco}, {Sacco}, {Smiljanic}, {Van Eck}, \&
  {Walton}}]{2012Msngr.147...25G}
{Gilmore}, G., {Randich}, S., {Asplund}, M., {et~al.} 2012, The Messenger, 147,
  25

\bibitem[{{Girardi}(1999)}]{1999MNRAS.308..818G}
{Girardi}, L. 1999, \mnras, 308, 818

\bibitem[{{Girardi} {et~al.}(2000){Girardi}, {Mermilliod}, \&
  {Carraro}}]{2000A&A...354..892G}
{Girardi}, L., {Mermilliod}, J.-C., \& {Carraro}, G. 2000, \aap, 354, 892

\bibitem[{{Gonzalez} {et~al.}(2009){Gonzalez}, {Zoccali}, {Monaco}, {Hill},
  {Cassisi}, {Minniti}, {Renzini}, {Barbuy}, {Ortolani}, \&
  {Gomez}}]{2009A&A...508..289G}
{Gonzalez}, O.~A., {Zoccali}, M., {Monaco}, L., {et~al.} 2009, \aap, 508, 289

\bibitem[{{Greenstein} \& {Richardson}(1951)}]{1951ApJ...113..536G}
{Greenstein}, J.~L. \& {Richardson}, R.~S. 1951, \apj, 113, 536

\bibitem[{{Grevesse} {et~al.}(2007){Grevesse}, {Asplund}, \&
  {Sauval}}]{2007SSRv..130..105G}
{Grevesse}, N., {Asplund}, M., \& {Sauval}, A.~J. 2007, \ssr, 130, 105

\bibitem[{{Hill} \& {Pasquini}(1999)}]{1999A&A...348L..21H}
{Hill}, V. \& {Pasquini}, L. 1999, \aap, 348, L21

\bibitem[{{Jasniewicz} {et~al.}(1999){Jasniewicz}, {Parthasarathy}, {de
  Laverny}, \& {Th{\'e}venin}}]{1999A&A...342..831J}
{Jasniewicz}, G., {Parthasarathy}, M., {de Laverny}, P., \& {Th{\'e}venin}, F.
  1999, \aap, 342, 831

\bibitem[{{Kirby} {et~al.}(2012){Kirby}, {Fu}, {Guhathakurta}, \&
  {Deng}}]{2012ApJ...752L..16K}
{Kirby}, E.~N., {Fu}, X., {Guhathakurta}, P., \& {Deng}, L. 2012, \apjl, 752,
  L16

\bibitem[{{Kirby} {et~al.}(2016){Kirby}, {Guhathakurta}, {Zhang}, {Hong},
  {Guo}, {Guo}, {Cohen}, \& {Cunha}}]{2016ApJ...819..135K}
{Kirby}, E.~N., {Guhathakurta}, P., {Zhang}, A.~J., {et~al.} 2016, \apj, 819,
  135

\bibitem[{{Kriskovics} {et~al.}(2014){Kriskovics}, {K{\H o}v{\'a}ri}, {Vida},
  {Granzer}, \& {Ol{\'a}h}}]{2014A&A...571A..74K}
{Kriskovics}, L., {K{\H o}v{\'a}ri}, Z., {Vida}, K., {Granzer}, T., \&
  {Ol{\'a}h}, K. 2014, \aap, 571, A74

\bibitem[{{Kumar} {et~al.}(2011){Kumar}, {Reddy}, \&
  {Lambert}}]{2011ApJ...730L..12K}
{Kumar}, Y.~B., {Reddy}, B.~E., \& {Lambert}, D.~L. 2011, \apjl, 730, L12

\bibitem[{{Lagarde} {et~al.}(2012){Lagarde}, {Decressin}, {Charbonnel},
  {Eggenberger}, {Ekstr{\"o}m}, \& {Palacios}}]{2012A&A...543A.108L}
{Lagarde}, N., {Decressin}, T., {Charbonnel}, C., {et~al.} 2012, \aap, 543,
  A108

\bibitem[{{Lattanzio} {et~al.}(2015){Lattanzio}, {Siess}, {Church}, {Angelou},
  {Stancliffe}, {Doherty}, {Stephen}, \& {Campbell}}]{2015MNRAS.446.2673L}
{Lattanzio}, J.~C., {Siess}, L., {Church}, R.~P., {et~al.} 2015, \mnras, 446,
  2673

\bibitem[{{L{\`e}bre} {et~al.}(1999){L{\`e}bre}, {de Laverny}, {de Medeiros},
  {Charbonnel}, \& {da Silva}}]{1999A&A...345..936L}
{L{\`e}bre}, A., {de Laverny}, P., {de Medeiros}, J.~R., {Charbonnel}, C., \&
  {da Silva}, L. 1999, \aap, 345, 936

\bibitem[{{L{\`e}bre} {et~al.}(2006){L{\`e}bre}, {de Laverny}, {Do Nascimento},
  \& {de Medeiros}}]{2006A&A...450.1173L}
{L{\`e}bre}, A., {de Laverny}, P., {Do Nascimento}, Jr., J.~D., \& {de
  Medeiros}, J.~R. 2006, \aap, 450, 1173

\bibitem[{{L{\`e}bre} {et~al.}(2009){L{\`e}bre}, {Palacios}, {Do Nascimento},
  {Konstantinova-Antova}, {Kolev}, {Auri{\`e}re}, {de Laverny}, \& {de
  Medeiros}}]{2009A&A...504.1011L}
{L{\`e}bre}, A., {Palacios}, A., {Do Nascimento}, Jr., J.~D., {et~al.} 2009,
  \aap, 504, 1011

\bibitem[{{Lind} {et~al.}(2009){Lind}, {Asplund}, \&
  {Barklem}}]{2009A&A...503..541L}
{Lind}, K., {Asplund}, M., \& {Barklem}, P.~S. 2009, \aap, 503, 541

\bibitem[{{Maeder} {et~al.}(2013){Maeder}, {Meynet}, {Lagarde}, \&
  {Charbonnel}}]{2013A&A...553A...1M}
{Maeder}, A., {Meynet}, G., {Lagarde}, N., \& {Charbonnel}, C. 2013, \aap, 553,
  A1

\bibitem[{{Martell} \& {Shetrone}(2013)}]{2013MNRAS.430..611M}
{Martell}, S.~L. \& {Shetrone}, M.~D. 2013, \mnras, 430, 611

\bibitem[{{McSwain} \& {Gies}(2005)}]{2005ApJS..161..118M}
{McSwain}, M.~V. \& {Gies}, D.~R. 2005, \apjs, 161, 118

\bibitem[{{Melo} {et~al.}(2005){Melo}, {de Laverny}, {Santos}, {Israelian},
  {Randich}, {Do Nascimento}, \& {de Medeiros}}]{2005A&A...439..227M}
{Melo}, C.~H.~F., {de Laverny}, P., {Santos}, N.~C., {et~al.} 2005, \aap, 439,
  227

\bibitem[{{Moc{\'a}k} {et~al.}(2011){Moc{\'a}k}, {Meakin}, {M{\"u}ller}, \&
  {Siess}}]{2011ApJ...743...55M}
{Moc{\'a}k}, M., {Meakin}, C.~A., {M{\"u}ller}, E., \& {Siess}, L. 2011, \apj,
  743, 55

\bibitem[{{Monaco} {et~al.}(2014){Monaco}, {Boffin}, {Bonifacio}, {Villanova},
  {Carraro}, {Caffau}, {Steffen}, {Ahumada}, {Beletsky}, \&
  {Beccari}}]{2014A&A...564L...6M}
{Monaco}, L., {Boffin}, H.~M.~J., {Bonifacio}, P., {et~al.} 2014, \aap, 564, L6

\bibitem[{{Monaco} {et~al.}(2011){Monaco}, {Villanova}, {Moni Bidin},
  {Carraro}, {Geisler}, {Bonifacio}, {Gonzalez}, {Zoccali}, \&
  {Jilkova}}]{2011A&A...529A..90M}
{Monaco}, L., {Villanova}, S., {Moni Bidin}, C., {et~al.} 2011, \aap, 529, A90

\bibitem[{{North}(1993)}]{1993ASPC...44..577N}
{North}, P. 1993, in Astronomical Society of the Pacific Conference Series,
  Vol.~44, IAU Colloq. 138: Peculiar versus Normal Phenomena in A-type and
  Related Stars, ed. M.~M. {Dworetsky}, F.~{Castelli}, \& R.~{Faraggiana}, 577

\bibitem[{{Palacios} {et~al.}(2001){Palacios}, {Charbonnel}, \&
  {Forestini}}]{2001A&A...375L...9P}
{Palacios}, A., {Charbonnel}, C., \& {Forestini}, M. 2001, \aap, 375, L9

\bibitem[{{Palacios} {et~al.}(2003){Palacios}, {Talon}, {Charbonnel}, \&
  {Forestini}}]{2003A&A...399..603P}
{Palacios}, A., {Talon}, S., {Charbonnel}, C., \& {Forestini}, M. 2003, \aap,
  399, 603

\bibitem[{{Pasquini} {et~al.}(2014){Pasquini}, {Koch}, {Smiljanic},
  {Bonifacio}, \& {Modigliani}}]{2014A&A...563A...3P}
{Pasquini}, L., {Koch}, A., {Smiljanic}, R., {Bonifacio}, P., \& {Modigliani},
  A. 2014, \aap, 563, A3

\bibitem[{{Pasquini} {et~al.}(2001){Pasquini}, {Randich}, \&
  {Pallavicini}}]{2001A&A...374.1017P}
{Pasquini}, L., {Randich}, S., \& {Pallavicini}, R. 2001, \aap, 374, 1017

\bibitem[{{Pasquini} {et~al.}(2004){Pasquini}, {Randich}, {Zoccali}, {Hill},
  {Charbonnel}, \& {Nordstr{\"o}m}}]{2004A&A...424..951P}
{Pasquini}, L., {Randich}, S., {Zoccali}, M., {et~al.} 2004, \aap, 424, 951

\bibitem[{{Pereyra} {et~al.}(2006){Pereyra}, {Castilho}, \&
  {Magalh{\~a}es}}]{2006A&A...449..211P}
{Pereyra}, A., {Castilho}, B.~V., \& {Magalh{\~a}es}, A.~M. 2006, \aap, 449,
  211

\bibitem[{{Pilachowski} {et~al.}(2000){Pilachowski}, {Sneden}, {Kraft},
  {Harmer}, \& {Willmarth}}]{2000AJ....119.2895P}
{Pilachowski}, C.~A., {Sneden}, C., {Kraft}, R.~P., {Harmer}, D., \&
  {Willmarth}, D. 2000, \aj, 119, 2895

\bibitem[{{Platais} {et~al.}(2012){Platais}, {Melo}, {Quinn}, {Clem}, {de
  Mink}, {Dotter}, {Kozhurina-Platais}, {Latham}, \&
  {Bellini}}]{2012ApJ...751L...8P}
{Platais}, I., {Melo}, C., {Quinn}, S.~N., {et~al.} 2012, \apjl, 751, L8

\bibitem[{{Power} {et~al.}(2007){Power}, {Wade}, {Hanes}, {Aurier}, \&
  {Silvester}}]{2007pms..conf...89P}
{Power}, J., {Wade}, G.~A., {Hanes}, D.~A., {Aurier}, M., \& {Silvester}, J.
  2007, in Physics of Magnetic Stars, ed. I.~I. {Romanyuk}, D.~O.
  {Kudryavtsev}, O.~M. {Neizvestnaya}, \& V.~M. {Shapoval}, 89--97

\bibitem[{{Randich} \& {Gilmore}(2013)}]{2013Msngr.154...47R}
{Randich}, S. \& {Gilmore}, G. 2013, The Messenger, 154, 47

\bibitem[{{Rebull} {et~al.}(2015){Rebull}, {Carlberg}, {Gibbs}, {Deeb},
  {Larsen}, {Black}, {Altepeter}, {Bucksbee}, {Cashen}, {Clarke}, {Datta},
  {Hodgson}, \& {Lince}}]{2015AJ....150..123R}
{Rebull}, L.~M., {Carlberg}, J.~K., {Gibbs}, J.~C., {et~al.} 2015, \aj, 150,
  123

\bibitem[{{Reddy} {et~al.}(2002){Reddy}, {Lambert}, {Hrivnak}, \&
  {Bakker}}]{2002AJ....123.1993R}
{Reddy}, B.~E., {Lambert}, D.~L., {Hrivnak}, B.~J., \& {Bakker}, E.~J. 2002,
  \aj, 123, 1993

\bibitem[{{Ruchti} {et~al.}(2011){Ruchti}, {Fulbright}, {Wyse}, {Gilmore},
  {Grebel}, {Bienaym{\'e}}, {Bland-Hawthorn}, {Freeman}, {Gibson}, {Munari},
  {Navarro}, {Parker}, {Reid}, {Seabroke}, {Siebert}, {Siviero}, {Steinmetz},
  {Watson}, {Williams}, \& {Zwitter}}]{2011ApJ...743..107R}
{Ruchti}, G.~R., {Fulbright}, J.~P., {Wyse}, R.~F.~G., {et~al.} 2011, \apj,
  743, 107

\bibitem[{{Sacco} {et~al.}(2014){Sacco}, {Morbidelli}, {Franciosini},
  {Maiorca}, {Randich}, {Modigliani}, {Gilmore}, {Asplund}, {Binney},
  {Bonifacio}, {Drew}, {Feltzing}, {Ferguson}, {Jeffries}, {Micela},
  {Negueruela}, {Prusti}, {Rix}, {Vallenari}, {Alfaro}, {Allende Prieto},
  {Babusiaux}, {Bensby}, {Blomme}, {Bragaglia}, {Flaccomio}, {Francois},
  {Hambly}, {Irwin}, {Koposov}, {Korn}, {Lanzafame}, {Pancino}, {Recio-Blanco},
  {Smiljanic}, {Van Eck}, {Walton}, {Bergemann}, {Costado}, {de Laverny},
  {Heiter}, {Hill}, {Hourihane}, {Jackson}, {Jofre}, {Lewis}, {Lind}, {Lardo},
  {Magrini}, {Masseron}, {Prisinzano}, \& {Worley}}]{2014A&A...565A.113S}
{Sacco}, G.~G., {Morbidelli}, L., {Franciosini}, E., {et~al.} 2014, \aap, 565,
  A113

\bibitem[{{Sackmann} \& {Boothroyd}(1999)}]{1999ApJ...510..217S}
{Sackmann}, I.-J. \& {Boothroyd}, A.~I. 1999, \apj, 510, 217

\bibitem[{{Siess} \& {Livio}(1999)}]{1999MNRAS.308.1133S}
{Siess}, L. \& {Livio}, M. 1999, \mnras, 308, 1133

\bibitem[{{Silva Aguirre} {et~al.}(2014){Silva Aguirre}, {Ruchti}, {Hekker},
  {Cassisi}, {Christensen-Dalsgaard}, {Datta}, {Jendreieck}, {Jessen-Hansen},
  {Mazumdar}, {Mosser}, {Stello}, {Beck}, \& {de Ridder}}]{2014ApJ...784L..16S}
{Silva Aguirre}, V., {Ruchti}, G.~R., {Hekker}, S., {et~al.} 2014, \apjl, 784,
  L16

\bibitem[{{Smiljanic} {et~al.}(2014){Smiljanic}, {Korn}, {Bergemann}, {Frasca},
  {Magrini}, {Masseron}, {Pancino}, {Ruchti}, {San Roman}, {Sbordone}, {Sousa},
  {Tabernero}, {Tautvai{\v s}ien{\.e}}, {Valentini}, {Weber}, {Worley},
  {Adibekyan}, {Allende Prieto}, {Barisevi{\v c}ius}, {Biazzo},
  {Blanco-Cuaresma}, {Bonifacio}, {Bragaglia}, {Caffau}, {Cantat-Gaudin},
  {Chorniy}, {de Laverny}, {Delgado-Mena}, {Donati}, {Duffau}, {Franciosini},
  {Friel}, {Geisler}, {Gonz{\'a}lez Hern{\'a}ndez}, {Gruyters}, {Guiglion},
  {Hansen}, {Heiter}, {Hill}, {Jacobson}, {Jofre}, {J{\"o}nsson}, {Lanzafame},
  {Lardo}, {Ludwig}, {Maiorca}, {Mikolaitis}, {Montes}, {Morel}, {Mucciarelli},
  {Mu{\~n}oz}, {Nordlander}, {Pasquini}, {Puzeras}, {Recio-Blanco}, {Ryde},
  {Sacco}, {Santos}, {Serenelli}, {Sordo}, {Soubiran}, {Spina}, {Steffen},
  {Vallenari}, {Van Eck}, {Villanova}, {Gilmore}, {Randich}, {Asplund},
  {Binney}, {Drew}, {Feltzing}, {Ferguson}, {Jeffries}, {Micela}, {Negueruela},
  {Prusti}, {Rix}, {Alfaro}, {Babusiaux}, {Bensby}, {Blomme}, {Flaccomio},
  {Fran{\c c}ois}, {Irwin}, {Koposov}, {Walton}, {Bayo}, {Carraro}, {Costado},
  {Damiani}, {Edvardsson}, {Hourihane}, {Jackson}, {Lewis}, {Lind}, {Marconi},
  {Martayan}, {Monaco}, {Morbidelli}, {Prisinzano}, \&
  {Zaggia}}]{2014A&A...570A.122S}
{Smiljanic}, R., {Korn}, A.~J., {Bergemann}, M., {et~al.} 2014, \aap, 570, A122

\bibitem[{{Stello} {et~al.}(2016){Stello}, {Cantiello}, {Fuller}, {Huber},
  {Garc{\'{\i}}a}, {Bedding}, {Bildsten}, \& {Aguirre}}]{2016Natur.529..364S}
{Stello}, D., {Cantiello}, M., {Fuller}, J., {et~al.} 2016, \nat, 529, 364

\bibitem[{{Strassmeier} {et~al.}(2015){Strassmeier}, {Carroll}, {Weber}, \&
  {Granzer}}]{2015A&A...574A..31S}
{Strassmeier}, K.~G., {Carroll}, T.~A., {Weber}, M., \& {Granzer}, T. 2015,
  \aap, 574, A31

\bibitem[{{Tautvai{\v s}ien{\.e}} {et~al.}(2015){Tautvai{\v s}ien{\.e}},
  {Drazdauskas}, {Mikolaitis}, {Barisevi{\v c}ius}, {Puzeras}, {Stonkut{\.e}},
  {Chorniy}, {Magrini}, {Romano}, {Smiljanic}, {Bragaglia}, {Carraro}, {Friel},
  {Morel}, {Pancino}, {Donati}, {Jim{\'e}nez-Esteban}, {Gilmore}, {Randich},
  {Jeffries}, {Vallenari}, {Bensby}, {Flaccomio}, {Recio-Blanco}, {Costado},
  {Hill}, {Jofr{\'e}}, {Lardo}, {de Laverny}, {Masseron}, {Moribelli}, {Sousa},
  \& {Zaggia}}]{2015A&A...573A..55T}
{Tautvai{\v s}ien{\.e}}, G., {Drazdauskas}, A., {Mikolaitis}, {\v S}., {et~al.}
  2015, \aap, 573, A55

\bibitem[{{Wallerstein} \& {Sneden}(1982)}]{1982ApJ...255..577W}
{Wallerstein}, G. \& {Sneden}, C. 1982, \apj, 255, 577

\end{thebibliography}

\Online
\begin{table*}
 \caption[]{\label{tab:obs} Observational data for the member stars of Trumpler 20.}
\centering
\begin{tabular}{cccccccr}
\hline
\hline
Star ID & Gaia-ESO ID & R.A. & DEC & $V$ & ($B-V$) & RV & S/N \\
 & & deg (J2000) & deg (J2000) & mag & mag & km s$^{-1}$ & \\
\hline
63    & 12385807-6030286   &  189.7420   &  $-$60.5079  &  13.60 &  1.59 &  $-$40.81 &  109  \\
129   & 12400109-6031395   &  190.0046   &  $-$60.5276  &  14.72 &  1.42 &  $-$40.04 &  58   \\
203   & 12393740-6032568   &  189.9059   &  $-$60.5491  &  14.89 &  1.35 &  $-$40.18 &  43   \\
227   & 12394385-6033165   &  189.9328   &  $-$60.5546  &  14.61 &  1.34 &  $-$40.50 &  50   \\
246   & 12394897-6033282   &  189.9541   &  $-$60.5578  &  14.55 &  1.34 &  $-$39.07 &  44   \\
287   & 12394688-6033540   &  189.9454   &  $-$60.5650  &  14.80 &  1.35 &  $-$40.48 &  53   \\
292   & 12390409-6034001   &  189.7671   &  $-$60.5667  &  13.41 &  1.55 &  $-$40.08 &  98   \\
340   & 12391577-6034406   &  189.8157   &  $-$60.5779  &  14.74 &  1.35 &  $-$40.21 &  49   \\
346   & 12394418-6034410   &  189.9341   &  $-$60.5781  &  14.71 &  1.37 &  $-$40.50 &  50   \\
399   & 12395973-6035072   &  189.9990   &  $-$60.5853  &  14.55 &  1.40 &  $-$41.74 &  78   \\
429   & 12400116-6035218   &  190.0048   &  $-$60.5894  &  14.54 &  1.37 &  $-$39.27 &  59   \\
430   & 12395566-6035233   &  189.9820   &  $-$60.5898  &  15.25 &  1.00 &  $-$41.79 &  35   \\
468   & 12400754-6035445   &  190.0315   &  $-$60.5957  &  13.47 &  1.61 &  $-$39.86 &  83   \\
505   & 12392698-6036053   &  189.8625   &  $-$60.6015  &  14.59 &  1.42 &  $-$40.31 &  66   \\
542   & 12391200-6036322   &  189.8000   &  $-$60.6089  &  14.69 &  1.33 &  $-$40.58 &  60   \\
582   & 12391113-6036528   &  189.7964   &  $-$60.6146  &  14.92 &  1.36 &  $-$40.76 &  49   \\
591   & 12400449-6036566   &  190.0188   &  $-$60.6157  &  13.62 &  1.64 &  $-$40.73 &  88   \\
638   & 12395554-6037268   &  189.9815   &  $-$60.6241  &  14.64 &  1.39 &  $-$40.07 &  83   \\
679   & 12402227-6037419   &  190.0929   &  $-$60.6283  &  14.74 &  1.42 &  $-$41.13 &  47   \\
724   & 12390709-6038056   &  189.7796   &  $-$60.6349  &  14.92 &  1.31 &  $-$38.38 &  46   \\
768   & 12394514-6038258   &  189.9382   &  $-$60.6405  &  14.80 &  1.38 &  $-$40.68 &  56   \\
770   & 12392584-6038279   &  189.8577   &  $-$60.6411  &  14.93 &  1.34 &  $-$42.97 &  36   \\
781   & 12394475-6038339   &  189.9365   &  $-$60.6427  &  14.62 &  1.40 &  $-$37.89 &  52   \\
787   & 12395424-6038370   &  189.9761   &  $-$60.6436  &  14.61 &  1.39 &  $-$42.48 &  65   \\
791   & 12394596-6038389   &  189.9415   &  $-$60.6441  &  14.54 &  1.38 &  $-$38.96 &  65   \\
794   & 12391002-6038402   &  189.7918   &  $-$60.6445  &  13.57 &  1.57 &  $-$40.60 &  87   \\
795   & 12394742-6038411   &  189.9476   &  $-$60.6447  &  14.72 &  1.36 &  $-$39.30 &  46   \\
827   & 12395654-6039012   &  189.9856   &  $-$60.6503  &  14.78 &  1.38 &  $-$38.23 &  57   \\
835   & 12393781-6039051   &  189.9076   &  $-$60.6514  &  14.49 &  1.35 &  $-$39.71 &  45   \\
858   & 12394307-6039193   &  189.9295   &  $-$60.6554  &  14.63 &  1.42 &  $-$40.82 &  54   \\
885   & 12395711-6039335   &  189.9880   &  $-$60.6593  &  14.64 &  1.37 &  $-$41.01 &  62   \\
894   & 12393131-6039423   &  189.8805   &  $-$60.6617  &  14.77 &  1.34 &  $-$35.29 &  56   \\
911   & 12400259-6039545   &  190.0108   &  $-$60.6651  &  13.74 &  1.61 &  $-$40.41 &  106  \\
923   & 12394121-6040040   &  189.9217   &  $-$60.6678  &  14.84 &  1.37 &  $-$41.01 &  56   \\
950   & 12392636-6040217   &  189.8599   &  $-$60.6727  &  14.74 &  1.35 &  $-$40.67 &  42   \\
1008  & 12394715-6040584   &  189.9465   &  $-$60.6829  &  13.85 &  1.57 &  $-$39.41 &  78  \\
1010  & 12394049-6041006   &  189.9188   &  $-$60.6835  &  14.56 &  1.37 &  $-$41.99 &  48  \\
1044  & 12400278-6041192   &  190.0116   &  $-$60.6887  &  14.95 &  1.38 &  $-$38.75 &  40  \\
1082  & 12390478-6041475   &  189.7699   &  $-$60.6965  &  14.59 &  1.32 &  $-$40.15 &  52  \\
2690  & 12383657-6045300   &  189.6525   &  $-$60.7583  &  14.57 &  1.44 &  $-$41.62 &  27  \\
2730  & 12383597-6045242   &  189.6498   &  $-$60.7568  &  14.94 &  1.38 &  $-$40.08 &  53  \\
3470  & 12402478-6043103   &  190.1033   &  $-$60.7195  &  13.92 &  1.60 &  $-$39.84 &  57  \\
\hline
\end{tabular}
\tablefoot{The star ID is taken from \citet{2005ApJS..161..118M}. The $V$ magnitude and ($B-V$) color have been corrected from differential reddening by \citet{2014A&A...561A..94D}. The radial velocities and signal-to-noise values were determined from the bluer part of the UVES spectrum as described in \citet{2014A&A...565A.113S}.}
\end{table*}

\begin{table*}
 \caption[]{\label{tab:par} Atmospheric parameters and lithium abundances for the member stars of Trumpler 20.}
\centering
\begin{tabular}{ccccccccccccc}
\hline
\hline
Star ID & T$_{\rm eff}$ & $\sigma$ & $\log~g$ & $\sigma$ & [Fe/H] & $\sigma$ & $\xi$ & $\sigma$ & A(Li)$_{\rm LTE}$ & $\sigma$ & Flag & A(Li)$_{\rm non-LTE}$\\
            &   (K) & (K) & (dex) & (dex) & (dex) & (dex) & km s$^{-1}$ & km s$^{-1}$ & (dex) & (dex) & \\
\hline
63   &     4551  &    133  &    2.18  &    0.29  &     0.09  &    0.11  &    1.45  &    0.06  &    0.50  &    0.38  &   lim.  &   0.75     \\
129  &     4888  &    116  &    2.85  &    0.23  &     0.13  &    0.10  &    1.39  &    0.04  &    0.85  &    0.05  &   lim.  &   1.02     \\
203  &     5031  &    120  &    3.00  &    0.22  &     0.14  &    0.10  &    1.40  &    0.10  &    1.07  &    0.14  &   lim.  &   1.22     \\
227  &     5004  &    113  &    2.87  &    0.23  &     0.09  &    0.10  &    1.59  &    0.06  &    1.04  &    0.18  &   lim.  &   1.20     \\
246  &     4947  &    114  &    2.81  &    0.22  &     0.10  &    0.09  &    1.50  &    0.09  &    1.03  &    0.18  &   lim.  &   1.21     \\
287  &     4961  &    116  &    2.90  &    0.23  &     0.14  &    0.09  &    1.38  &    0.06  &    0.87  &    0.12  &   lim.  &   1.03     \\
292  &     4406  &    114  &    2.05  &    0.24  &     0.00  &    0.10  &    1.55  &    0.09  &    0.28  &    0.34  &   lim.  &   0.60     \\
340  &     4851  &    118  &    2.88  &    0.23  &     0.02  &    0.10  &    1.36  &    0.06  &    1.37  &    0.21  &   det.  &   1.54     \\
346  &     4963  &    118  &    2.81  &    0.23  &     0.15  &    0.10  &    1.47  &    0.13  &    0.92  &    0.14  &   lim.  &   1.08     \\
399  &     4876  &    113  &    2.76  &    0.22  &     0.10  &    0.10  &    1.41  &    0.06  &    0.77  &    0.22  &   lim.  &   0.95     \\
429  &     4887  &    122  &    2.77  &    0.22  &     0.10  &    0.10  &    1.38  &    0.05  &    0.70  &    0.10  &   lim.  &   0.87     \\
430  &     5992  &    125  &    3.79  &    0.25  &     0.20  &    0.10  &    1.64  &    0.21  &    2.83  &    0.21  &   det.  &   2.81     \\
468  &     4435  &    115  &    2.10  &    0.23  &     0.06  &    0.10  &    1.52  &    0.05  &    0.34  &    0.31  &   lim.  &   0.65     \\
505  &     4834  &    120  &    2.76  &    0.25  &     0.11  &    0.10  &    1.39  &    0.10  &    1.07  &    0.21  &   det.  &   1.25     \\ 
542  &     4939  &    112  &    2.83  &    0.23  &     0.15  &    0.10  &    1.33  &    0.05  &    0.93  &    0.20  &   lim.  &   1.09     \\
582  &     4967  &    115  &    2.93  &    0.23  &     0.18  &    0.10  &    1.39  &    0.11  &    0.97  &    0.15  &   lim.  &   1.13     \\
591  &     4412  &    132  &    2.14  &    0.25  &     0.00  &    0.10  &    1.59  &    0.23  &    1.32  &    0.21  &   det.  &   1.60     \\
638  &     4900  &    112  &    2.79  &    0.22  &     0.13  &    0.10  &    1.38  &    0.05  &    0.74  &    0.20  &   lim.  &   0.91     \\
679  &     4936  &    121  &    2.77  &    0.23  &     0.12  &    0.10  &    1.50  &    0.10  &    0.92  &    0.15  &   lim.  &   1.08     \\
724  &     5026  &    120  &    3.14  &    0.23  &     0.10  &    0.10  &    1.27  &    0.04  &    1.01  &    0.13  &   lim.  &   1.15     \\
768  &     4928  &    119  &    2.85  &    0.23  &     0.12  &    0.10  &    1.36  &    0.05  &    0.85  &    0.15  &   lim.  &   1.01     \\
770  &     5035  &    119  &    3.14  &    0.25  &     0.12  &    0.10  &    1.28  &    0.07  &    1.15  &    0.14  &   lim.  &   1.29     \\ 
781  &     4882  &    118  &    2.77  &    0.23  &     0.12  &    0.10  &    1.48  &    0.09  &    0.88  &    0.18  &   lim.  &   1.05     \\ 
787  &     4913  &    109  &    2.80  &    0.23  &     0.14  &    0.10  &    1.42  &    0.04  &    0.90  &    0.22  &   lim.  &   1.07     \\
791  &     4889  &    122  &    2.75  &    0.24  &     0.10  &    0.11  &    1.42  &    0.06  &    0.70  &    0.13  &   lim.  &   0.87     \\
794  &     4471  &    118  &    2.14  &    0.23  &     0.01  &    0.10  &    1.56  &    0.17  &    0.43  &    0.05  &   lim.  &   0.74     \\
795  &     4924  &    117  &    2.77  &    0.23  &     0.08  &    0.10  &    1.50  &    0.07  &    0.96  &    0.17  &   lim.  &   1.13     \\
827  &     4932  &    122  &    2.83  &    0.25  &     0.16  &    0.10  &    1.39  &    0.05  &    0.89  &    0.17  &   lim.  &   1.05     \\
835  &     4935  &    128  &    2.79  &    0.22  &     0.12  &    0.11  &    1.44  &    0.14  &    0.87  &    0.11  &   lim.  &   1.03     \\
858  &     4880  &    117  &    2.76  &    0.23  &     0.12  &    0.09  &    1.34  &    0.05  &    0.69  &    0.11  &   lim.  &   0.87     \\
885  &     4964  &    116  &    2.83  &    0.23  &     0.14  &    0.09  &    1.44  &    0.06  &    0.93  &    0.05  &   lim.  &   1.09     \\
894  &     4988  &    114  &    3.04  &    0.22  &     0.13  &    0.10  &    1.37  &    0.03  &    0.95  &    0.16  &   lim.  &   1.10     \\
911  &     4521  &    115  &    2.20  &    0.22  &     0.03  &    0.09  &    1.47  &    0.15  &    0.50  &    0.39  &   lim.  &   0.76     \\
923  &     4989  &    117  &    2.94  &    0.25  &     0.13  &    0.09  &    1.41  &    0.03  &    0.96  &    0.17  &   lim.  &   1.11     \\
950  &     4953  &    121  &    2.87  &    0.23  &     0.08  &    0.10  &    1.47  &    0.23  &    1.06  &    0.22  &   det.  &   1.22     \\ 
1008  &    4586  &    123  &    2.29  &    0.24  &     0.05  &    0.10  &    1.49  &    0.08  &    0.43  &    0.05  &   lim.  &   0.67     \\
1010  &    4926  &    120  &    2.82  &    0.24  &     0.10  &    0.11  &    1.40  &    0.10  &    0.94  &    0.16  &   lim.  &   1.11     \\
1044  &    4951  &    117  &    2.94  &    0.22  &     0.12  &    0.09  &    1.34  &    0.10  &    0.97  &    0.05  &   lim.  &   1.13     \\
1082  &    5003  &    116  &    2.95  &    0.22  &     0.18  &    0.09  &    1.47  &    0.07  &    1.00  &    0.16  &   lim.  &   1.15     \\
2690  &    4689  &    128  &    2.52  &    0.25  &    $-$0.01  &    0.12  &    1.37  &    0.06  &    0.68  &    0.05  &   lim.  &   0.90     \\
2730  &    5003  &    117  &    2.97  &    0.24  &     0.14  &    0.09  &    1.51  &    0.08  &    0.99  &    0.17  &   lim.  &   1.15     \\
3470  &    4590  &    123  &    2.28  &    0.22  &     0.03  &    0.11  &    1.48  &    0.10  &    0.51  &    0.18  &   lim.  &   0.75     \\
\hline
\end{tabular}
\tablefoot{The star ID is taken from \citet{2005ApJS..161..118M}. For the metallicity, we adopt A(Fe)$_{\odot}$ = 7.45 as solar reference \citep{2007SSRv..130..105G}. The flag indicates if the Li abundance is a detection or an upper limit.}
\end{table*}

\end{document}